\documentclass[prd,aps,preprintnumbers,amssymb]{revtex4}
\usepackage{graphicx}
\usepackage{epsf}
\usepackage{amsmath}
\usepackage{epstopdf}
\usepackage{multirow}
\usepackage{bm}
\def\e3p{$\eta \rightarrow 3 \pi$}

\begin{document}
\title{%
\hfill{\normalsize\vbox{%
\hbox{}
 }}\\
{The infrared or ultraviolet scale for a gauge theory }}

\author{Renata Jora
$^{\it \bf a}$~\footnote[1]{Email:
 rjora@theory.nipne.ro}}

\affiliation{$^{\bf \it a}$ National Institute of Physics and Nuclear Engineering PO Box MG-6, Bucharest-Magurele, Romania}

\date{\today}

\begin{abstract}
We propose a conjecture that leads to the determination of the intrinsic scale (the scale at which the coupling constant diverges) for any gauge theory by comparison to another theory for which there is phenomenological information, provided that the two theories are placed in the same space-time volume.  This conjecture is verified independently for $U(1)_{em}$ and for $QCD$.  In this framework and from this perspective we show that the group $SO(10)$ is the only viable candidate among the  supersymmetric GUT groups $SU(5)$, $SO(10)$ and $E_6$.
\end{abstract}

\maketitle

\section{Introduction}

The main properties of a  gauge theory are intrinsically related to the behavior of the theory in the infrared or ultraviolet regime. For non-abelian gauge theory the largeness of the  coupling constant in the infrared regime is associated to confinement and to the formation of the bound states; on the contrary the smallness of the coupling constant in the ultraviolet regime is called asymptotic freedom. For an abelian gauge theory this behavior is reversed; the coupling constant is large in the ultraviolet regime and small in the infrared. In many situations although of the same essence abelian and non-abelian gauge theories are treated distinctly and have different characteristics.

For a theory confining in the infrared we associate a definite scale $\Lambda$ which is the scale where the coupling constant diverges. If this scale is defined at one loop is definitely renormalization scheme independent, if it is defined at many loops is renormalization scheme dependent. For supersymmetric gauge theories there are multiple scales also but the clearest choice is the scale where the holomorphic coupling diverges. For a theory that it is strong in the ultraviolet  there is an ultraviolet scale no matter how large where the  coupling constant diverges.

For example for QCD with $N_f$ flavors one can define at one loop:
\begin{eqnarray}
\frac{8\pi^2}{g^2(\mu)}=[\frac{11}{3}N-\frac{2}{3}N_f]\ln(\frac{\mu}{\Lambda_{QCD}}).
\label{defl877}
\end{eqnarray}

For supersymmetric QCD with $N_f$ flavors in the fundamental and antifundamental representation the importance of $\Lambda$ is even more undoubtable as $\Lambda$ as  the theory is saturated at one loop:
\begin{eqnarray}
\frac{8\pi^2}{g_h^2(\mu)}=[3N-N_f]\ln(\frac{\mu}{\Lambda_{SQCD}}).
\label{holmpr8867}
\end{eqnarray}

It seems that $\Lambda$ however defined is very useful for describing a gauge theory and for defining an intrinsic scale of it. Unfortunately we know $\Lambda$ at least approximately only for theories for which we have phenomenological information. For example in QCD we know  the range for $\Lambda$, $0.2\leq \Lambda_{QCD}\leq0.36$ GeV. We have less information about the other theories.

The question that we will pose and answer at least partially in this paper is if and how one can determine $\Lambda$ in a gauge theory be that abelian or non-abelian, large or small in terms of only the beta function and anomalous dimensions in the theory and the volume of the space time associated with that theory. It will turn out that there are cases in which exact information can be obtained and cases in which although some knowledge might be extracted it depends in such measure on the parameters of the theory that it becomes useless.

Section II contains a general discussion and the set-up. In section II we will give clear examples and also make some predictions. The last section is dedicated to Conclusions.

\section{The $\Lambda$ conjecture}

Consider the running of the coupling constant with the scale.  The coupling constant depends not only on the beta function and on the intrinsic scale of it but also on all mass scales and coupling constants in the theories. The most dramatic change in the coupling constant behavior however is associated with the mass scales. Below this scale some matter fields decouple which leads to different beta function. For example for the supersymmetric QCD the introduction of masses for the $N_f$ matter fields leads to the decoupling of these below the scale $m$ according to the following procedure \cite{Seiberg}.  The coupling constants of the initial theory and for the decoupled one must be the same:
\begin{eqnarray}
\frac{8\pi^2}{g_h^2}=[3N-N_f]\ln(\frac{m}{\Lambda_1})=3N\ln(\frac{m}{\Lambda_2}),
\label{dec4553}
\end{eqnarray}
where $\Lambda_1$ and $\Lambda_2$ are the holomorphic scales for the two theories. This leads to:
\begin{eqnarray}
\Lambda_2=\Lambda_1^{\frac{3N-N_f}{3N}}m^{\frac{N_f}{3N}}.
\label{secdec45342}
\end{eqnarray}

It is evident that $\Lambda_2$ for the low energy theory is dependent not only on $\Lambda_1$ of the initial theory but also on the mass scale $m$.  The decoupled theory is governed not only by the infrared scale of the initial theory but also by all mass parameters of this. This is true in general. For a theory with spontaneous symmetry breaking like the standard model the emergent gauge groups are dependent on the mass scale in the theory or on the value of the condensates which are the combining results of all the couplings with mass dimension or not that appear in the theory.  It seems that no independent information can be extracted even approximate.

In general the scale dependent information of a theory that we have  is extracted from experimental knowledge therefore refers either to the electroweak group or QCD. $\Lambda$ for QCD with more flavors can be determined with precision from the beta function and from the value of $\alpha_s$ at $m_Z$ \cite{PDG}. In \cite{Deur} a detailed review of the strong coupling constant of QCD both in the perturbative and non-perturbative regime is given. There are many ways for determining $\alpha_S$ at the $m_Z$ scale from which we enumerate only a few: from hadronic reactions from lattice QCD studies, from $e^+e^-$ collisions, from $pp$ reactions, from light-front holographic QCD and even by fitting the parameters in supersymmetric gauge unificatiopn (\cite{Deur} and the references therein). The estimates for other gauge couplings in the standard model depend on the particular ultraviolet completion of the theory or if that theory is supersymmetric or not.

Considering all these facts facts it seems that too many variables enter in the determination of the infrared or ultraviolet scale of a theory  among which the behavior of the coupling constant and the dynamics of the Lagrangian are crucial. Therefore it appears that without phenomenological information no prediction can be made.

Next we will show that  there are cases where one can make definite predictions.  Let us consider a generic gauge theory, abelian or non-abelian. As main premise we consider that the theory is scale invariant at tree level. For concreteness let us assume the QCD Lagrangian:
\begin{eqnarray}
{\cal L}=-\frac{1}{4}F^{a\mu\nu}F^a_{\mu\nu}+\sum_f\bar{\Psi}_fi\gamma^{\mu}D_{\mu}\Psi_f,
\label{lagr45665}
\end{eqnarray}
where $D_{\mu}=\partial_{\mu}-igt^aA^a_{\mu}$ is the covariant derivative. We want to compare this theory with another in which matter fields or gauge fields are decoupled. In order to do that we must introduce masses and decouple at the corresponding scales. But masses break already at tree level the scale invariance and spoil our arguments. Assume that the volume of the space time is fixed finite or infinite and given by $V$. Then the only way one can introduce a scale invariant mass term for the gauge or matter of fields is to consider masses of order $V^{-1/4}$ as follows:
\begin{eqnarray}
&&cV^{-1/4}\bar{\Psi}_f\Psi_f
\nonumber\\
&&c^2V^{-1/2}A^a_{\mu}A^{a\mu},
\label{masvol8677}
\end{eqnarray}
where $c$ is the a dimensionless constant where in the case of a lattice is associated to the numbers of cells in the lattice $N$.

It is evident then that  at tree level the scale invariance is respected. At quantum level these mass terms will gain anomalous dimensions:
\begin{eqnarray}
&&cV^{-1/4}(1+\gamma_m)\bar{\Psi}_f\Psi_f
\nonumber\\
&&cV^{-1/2}[2+2\frac{\beta(g)}{g}]A^a_{\mu}A^{a\mu}.
\label{masvol867799}
\end{eqnarray}

Now we will make a leap and formulate the following conjecture: { \it For any scale invariant gauge theory the natural decoupling scale which relates one theory to another is the reduced scale (divided by $1$ GeV and thus dimensionless) $\frac{\mu}{1}=[\frac{cV^{-1/4}}{1}]^{(1+\gamma_m)}$ or $\frac{\mu}{1}=[\frac{cV^{-1/4}}{1}]^{(1+\frac{\beta(g)}{g})}$ specific to each theory. At this scale for the same space time volume the coupling constant for any gauge theory should be universal. Since theories may contain different matter multiplets with different anomalous dimensions the most reliable choice is however $\frac{\mu}{1}=[\frac{cV^{-1/4}}{1}]^{(1+\frac{\beta(g)}{g})}$. One does not need to consider specific decoupling patterns and even simply correlated matter and gauge content.
The nature of the gauge theory, abelian or non-abelian, supersymmetric or not is of no importance at least in first orders. Thus knowing the intrinsic scale of a theory allows the determination of the $\Lambda$ scale for any other gauge theory in the same space-time volume.}

Here we considered $\mu_0=1$ GeV as  the relevant infrared cut-off scale (can be taken the same as the renormalization scale). From the string theory and large dimensional compactification perspective an appropriate infrared cut-off scale would be $\mu_0=1-100$ TeV \cite{Hewett}, \cite{Cortes1}, \cite{Cortes2}. There are  few relevant scales that one might associate to the low energy QFT of the standard model; one of them is the electroweak scale as mentioned before. This scale is not appropriate from our point of view because we need to include the electroweak scale in the range of scales associated to the coupling constants. Next in order is $\mu_0=1$ GeV which is the approximate scale where the chiral symmetry breaking in QCD takes place and also where most low lying mesons and baryons are situated.  There are lower scales associated to the leptons but they are less important in association to the Planck scale.  We choose $\mu_0=1$ GeV as the relevant scale mostly  for its QCD related importance but one may find additional arguments as suggested in \cite{Hooman}, where it is envisioned that a dark matter candidate with mass around $1$ GeV might connect standard model with the Planck scale.

In what follows we will show that this conjecture works very well for all known theories and may also lead to consistent predictions. This will be done in the next section.

\section{Applications of the $\Lambda$ conjecture}

In this section we will first study in the conjecture discussed in section II those gauge theories  for which we have phenomenological information and then we will extrapolate to supersymmetric GUT theories to make some predictions. From the start we will eliminate from the discussion the theories that have mass scales associated to them and consider only those theories that run fully in the infrared or ultraviolet regime.  Therefore we will pick only QCD and $U(1)_{em}$. There are some difficulties associated to these from the point of view of matter degrees of freedom involved. Since the data that we have about them comes mainly from the electroweak scale we shall consider as the starting point the results for the coupling constants around that scale as taken from \cite{PDG}:
\begin{eqnarray}
&&\alpha(m_W)\approx\frac{1}{128}
\nonumber\\
&&\alpha_s(m_Z)\approx0.1181.
\label{const56748857}
\end{eqnarray}

Then right below the electroweak scale the theory has $5$ quarks and $3$ charged leptons that contribute and the beta functions at one loop for the two theories are:
\begin{eqnarray}
&&\beta(e)=\frac{e^3}{12\pi^2}\frac{60}{9}
\nonumber\\
&&\beta(g_s)=-\frac{g_s^3}{16\pi^2}\frac{23}{3}.
\label{thetwpbeta645}
\end{eqnarray}
One might consider to these beta functions two loop corrections for better accuracy. It turns out that at two loops there are electroweak corrections coming from other groups to both these beta functions \cite{Deur}, \cite{Vicini}. This would alter our conjecture such that we assume that the first order is a good approximation. The beta functions in Eq. (\ref{thetwpbeta645}) can be integrated out to lead to:
\begin{eqnarray}
&&\frac{1}{e^2(\mu_1)}-\frac{1}{e^2(m_W)}=\frac{60}{9}\frac{1}{6\pi^2}\ln(\frac{m_W}{\mu_1}).
\nonumber\\
&&\frac{1}{g_s^2(\mu_2)}-\frac{1}{g_s^2(m_Z)}=\frac{23}{3}\frac{1}{8\pi^2}\ln(\frac{\mu_2}{m_Z}).
\label{integr4536}
\end{eqnarray}
{\it Note that all the scales are divided by $1$ GeV and thus are rendered dimensionless. For the simplicity of the notation this will be considered implicit in what follows.}
We apply the conjecture stated in section II which requires that the two coupling constants $e$ and $g_s$ are equal at the scale $\mu_1=s^{1+\frac{\beta(e)}{e}}$ for QED and $\mu_2=s^{1+\frac{\beta(g_s)}{g_s}}$ for QCD, where $s$ is a scale associated to the space time volume which we consider the same for the two theories.  We know of only one scale associated to the real world space time structure which is the Planck scale. Therefore we will take $s=cV^{-1/4}=m_p\approx1.22091\times 10^{19}$ GeV.

We solve each equation in (\ref{integr4536}) individually with $\mu_1$ and $\mu_2$ given in the last paragraph to find two solutions for $U(1)_{em}$,
\begin{eqnarray}
&&e_1^2\approx0.176
\nonumber\\
&&e_2^2\approx20.406,
\label{sol76885}
\end{eqnarray}
and two solutions for QCD,
\begin{eqnarray}
g_{s1}^2\approx 0.224
\nonumber\\
g_{s2}^2\approx 21.510.
\label{sol891745}
\end{eqnarray}
Note the exceptional feature of these results. Even in the rough approximation in which we work $e_1^2$ and $g_{s1}^2$ are close in values and $e_2^2$ and $g_{s2}^2$ are even closer. So the requirement of the conjecture was obtained automatically by simply considering the right scale $s$. Of course it would be expected that appropriate two loop corrections would make the results even more accurate.

Then one can consider a general gauge theory with the beta function $\beta(g)$ and write for example (we consider for simplicity a non-abelian theory)  at one loop:
\begin{eqnarray}
\frac{8\pi^2}{g^2(\mu)}=\beta_0\ln(\frac{\mu}{\Lambda}),
\label{ex8675}
\end{eqnarray}
where $\Lambda$ is the infared scale of the theory. Taking $\mu=s^{1+\frac{\beta(g)}{g}}$ and requiring,
\begin{eqnarray}
g^2_s(s^{1+\frac{\beta(g_s)}{g_s}})=g^2(s^{1+\frac{\beta(g)}{g}}),
\label{cond6574664}
\end{eqnarray}
with $g^2$ extracted from Eq. (\ref{cond6574664}) leads to the determination of the scale $\Lambda$ of the unknown (or unresolved) theory situated in the same space-time volume $V$.

Next we will use our conjecture and the information we have up to this point to make a prediction with regard to the ultraviolet completion of the standard model. We will consider only supersymmetric extensions because it is known that these lead to more accurate gauge unification. Thus we will analyze  single large supersymmetric GUT groups: $SU(5)$, $SO(10)$ and $E_6$ \cite{Yanagida}-\cite{Veltman}.

First we will use the average value for the supersymmetrci gauge unification $\frac{g_g^2}{4\pi}=\frac{1}{24.3}$ and the scale at which this unfication takes place $\mu_G\approx 2\times 10^{16}$ GeV as taken from PDG \cite{PDG}.

For the unification group $SU(5)$ the matter is arranged in three generations of the antifundamental $5^*$ and antisymmetric $10$ representations. The two Higgs supermultiplets are situated in a $5$ and $5^*$. We shall consider that the gauge symmetry is broken through a dynamical mechanism by $5$ more Higgs supermultiplets in a $5$ and $5^*$ of the $SU(5)$ group. These correspond to the $5$ and $5^*$ representations of a second $SU(5)$ that provides the dynamical mechanism \cite{Jora}. The details of the gauge symmetry breaking however are not of great significance for our arguments. The corresponding beta function (which is an expansion of the NSVZ beta functions but any other expression could work as well) at two loops is given by:
\begin{eqnarray}
&&\beta(g_5)=-\frac{g_5^3}{16\pi^2}[a_1-a_2\frac{g_5^2}{8\pi^2}]
\nonumber\\
&&a_1=3
\nonumber\\
&&a_2=\frac{407}{5}.
\label{firstbetfunc645}
\end{eqnarray}

For the $SO(10)$ group there are three generations of matter in the spinor representation $16$. One can add a Higgs in the $10$ dimensional representation to break the gauge group but this modifies slightly the beta function. We choose to ignore for the time being the exact mechanism that breaks the gauge symmetry for both $SO(10)$ and $E_6$.  The corresponding beta function (again an expansion of the NSVZ beta function) is:
\begin{eqnarray}
&&\beta(g_{10})=-\frac{g_{10}^3}{16\pi^2}[x_1-x_2\frac{g_{10}^2}{8\pi^2}]
\nonumber\\
&&x_1=18
\nonumber\\
&&x_2=\frac{119}{2}.
\label{secgr7566488}
\end{eqnarray}

Finally we consider the exceptional group $E_6$. where the matter is placed in three generations of the $27$ dimensional representation. We choose the group factors as taken from \cite{Ritbergen} with $\eta=1$. The associated beta function (an expansion of the NSVZ beta function as before) is:
\begin{eqnarray}
&&\beta(g_6)=-\frac{g_6^3}{16\pi^2}[w_1-w_2\frac{g_6^2}{8\pi^2}]
\nonumber\\
&&w_1=54
\nonumber\\
&&w_2=600.
\label{thridgr6574}
\end{eqnarray}

We integrate the beta functions for each group as follows:
For $SU(5)$ we write:
\begin{eqnarray}
\int^{\mu_5}_{\mu_G}d[\frac{8\pi^2}{g_5^2}]\frac{1}{a_1-a_2\frac{g_5^2}{8\pi^2}}=\ln(\frac{\mu_5}{\mu_G}),
\label{intone756}
\end{eqnarray}
which is specified through the notation:
\begin{eqnarray}
Q(\mu_5)-Q(\mu_G)=\ln(\frac{\mu_5}{\mu_G}).
\label{notone75647}
\end{eqnarray}

For $SO(10)$ we write:
\begin{eqnarray}
\int^{\mu_{10}}_{\mu_G}d[\frac{8\pi^2}{g_{10}^2}]\frac{1}{x_1-x_2\frac{g_{10}^2}{8\pi^2}}=\ln(\frac{\mu_{10}}{\mu_G}),
\label{inttwo756}
\end{eqnarray}
which yields:
\begin{eqnarray}
H(\mu_{10})-H(\mu_G)=\ln(\frac{\mu_{10}}{\mu_G}).
\label{notone75647}
\end{eqnarray}

Finally $E_6$ leads to:
\begin{eqnarray}
\int^{\mu_{6}}_{\mu_G}d[\frac{8\pi^2}{g_{6}^2}]\frac{1}{w_1-w_2\frac{g_{6}^2}{8\pi^2}}=\ln(\frac{\mu_{6}}{\mu_G}),
\label{inttwo756}
\end{eqnarray}
and further on to:
\begin{eqnarray}
W(\mu_{6})-W(\mu_G)=\ln(\frac{\mu_{6}}{\mu_G}).
\label{notone75647}
\end{eqnarray}

Next we apply the conjecture by asking that at the scales:
\begin{eqnarray}
&&\mu_5=s^{1+\frac{\beta(g_5)}{g_5}}
\nonumber\\
&&\mu_{10}=s^{1+\frac{\beta(g_{10})}{g_{10}}}
\nonumber\\
&&\mu_6=s^{1+\frac{\beta(g_6)}{g_6}},
\label{finalexpr664553}
\end{eqnarray}
the coupling constants are approximately equal to $g_s$ at $\mu_1$.
For that we need to solve some equation that we choose to solve graphically. We introduce the functions:
\begin{eqnarray}
&&A(g_5)=Q(\mu_5)-Q(\mu_G)-\ln(\frac{\mu_5}{\mu_G}).
\nonumber\\
&&B(g_{10})=H(\mu_{10})-H(\mu_G)-\ln(\frac{\mu_{10}}{\mu_G})
\nonumber\\
&&C(g_6)=W(\mu_6)-W(\mu_G)-\ln(\frac{\mu_6}{\mu_G}).
\label{res645342}
\end{eqnarray}

\begin{figure}
\begin{center}
\epsfxsize = 10cm
\epsfbox{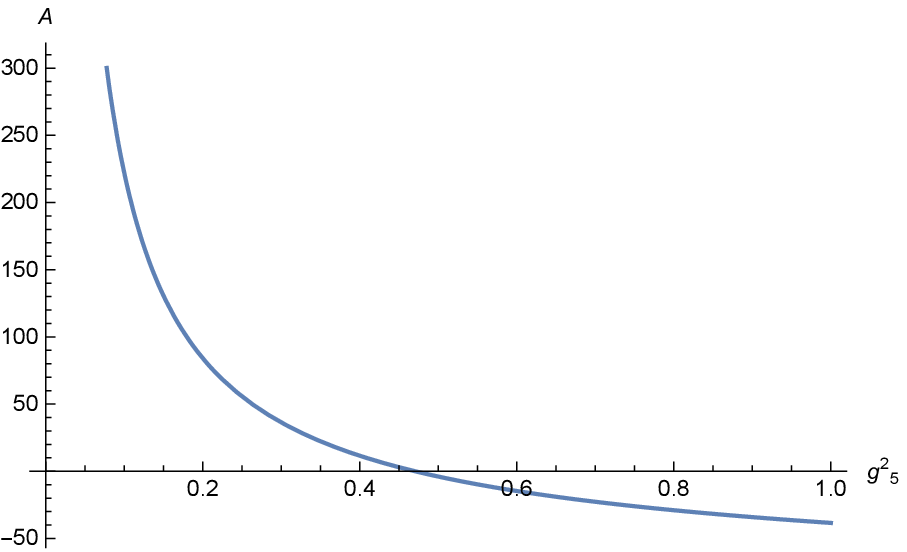}
\end{center}
\caption[]{%
Plot of $A$ in terms of $g_5^2$ in the range $(0.1)$.
}
\label{lambda1}
\end{figure}

\begin{figure}
\begin{center}
\epsfxsize = 10cm
\epsfbox{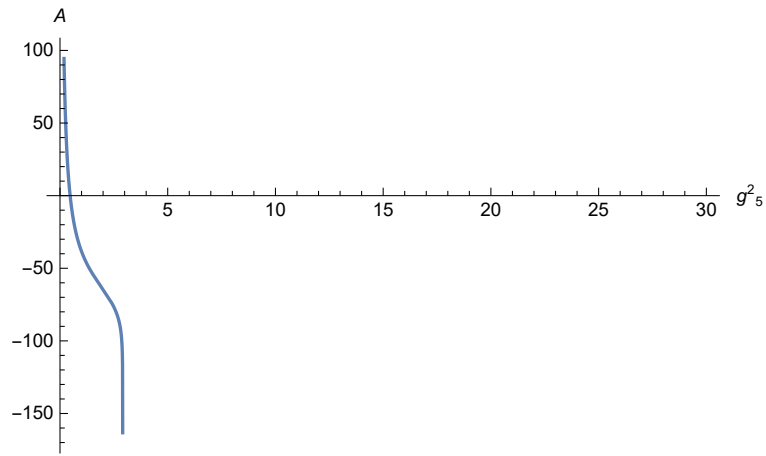}
\end{center}
\caption[]{%
Plot of $A$ in terms of $g_5^2$ in the range $(0,30)$.
}
\label{lambda2}
\end{figure}

We plot $A$ in the regions for the coupling constant squared $(0,1)$ and $(0,30)$ in Figs (\ref{lambda1}) and (\ref{lambda2}).

\begin{figure}
\begin{center}
\epsfxsize = 10cm
\epsfbox{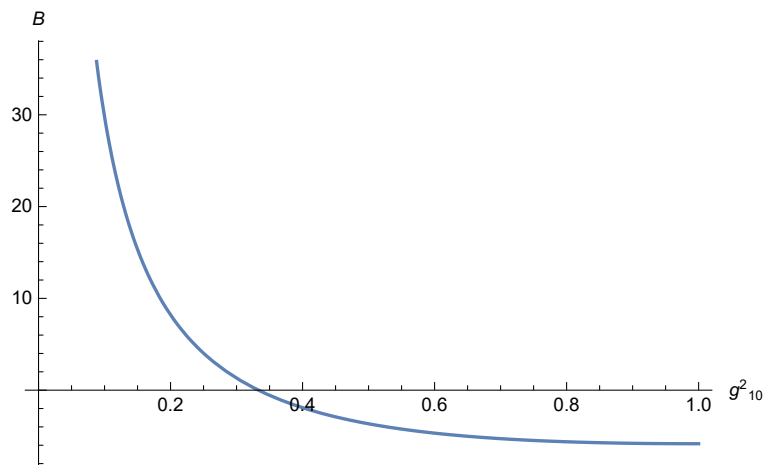}
\end{center}
\caption[]{%
Plot of $B$ in terms of $g_{10}^2$ in the range $(0,1)$.
}
\label{lambda3}
\end{figure}

\begin{figure}
\begin{center}
\epsfxsize = 10cm
\epsfbox{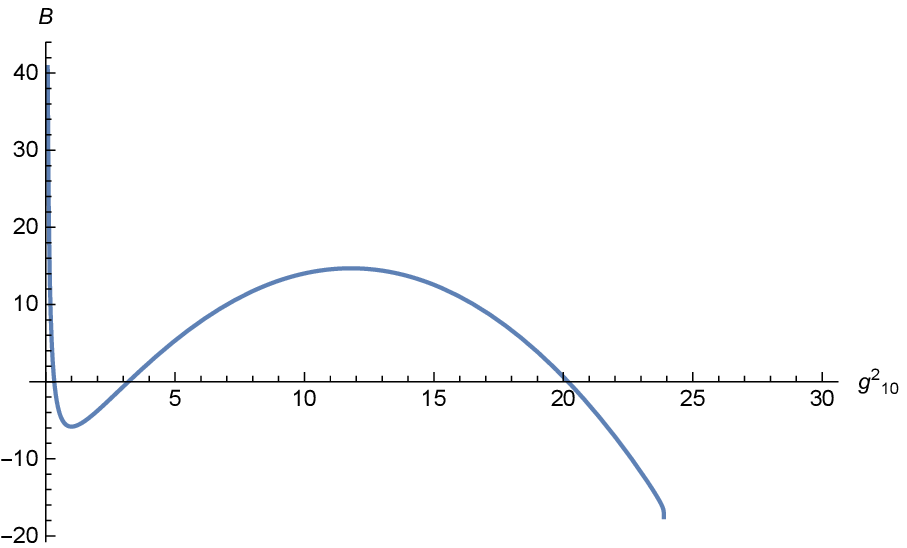}
\end{center}
\caption[]{%
Plot of $B$ in terms of $g_{10}^2$ in the range $(0,30)$.
}
\label{lambda4}
\end{figure}

We plot $B$ in the regions for the coupling constant squared $(0,1)$ and $(0,30)$ in Figs (\ref{lambda3}) and (\ref{lambda4}).


\begin{figure}
\begin{center}
\epsfxsize = 10cm
\epsfbox{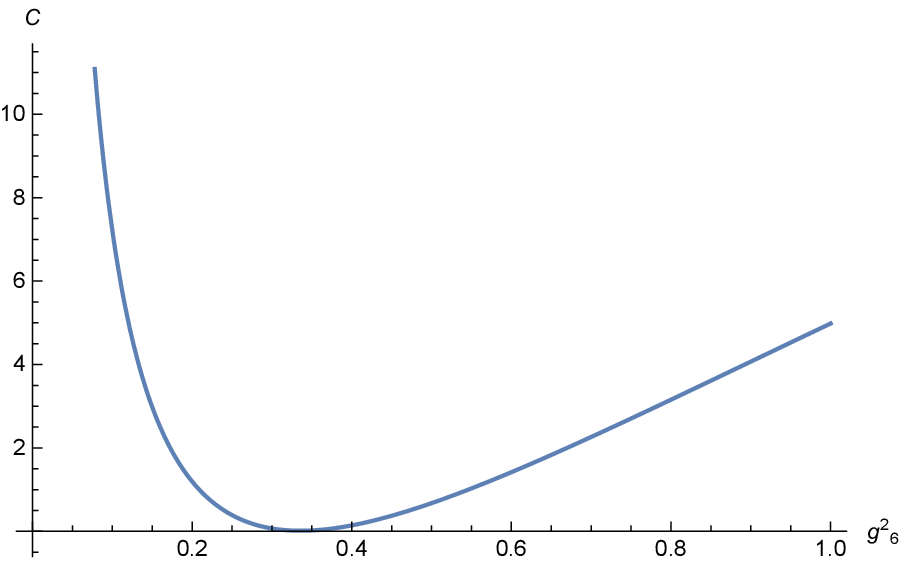}
\end{center}
\caption[]{%
Plot of $C$ in terms of $g_6^2$ in the range $(0,1)$.
}
\label{lambda5}
\end{figure}

\begin{figure}
\begin{center}
\epsfxsize = 10cm
\epsfbox{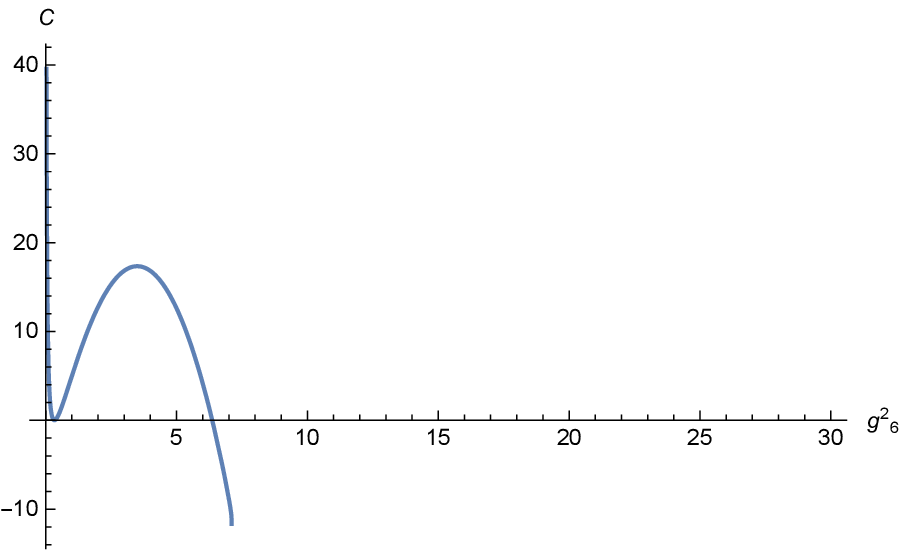}
\end{center}
\caption[]{%
Plot of $C$ in terms of $g_6^2$ in the range $(0,30)$.
}
\label{lambda6}
\end{figure}

We plot $C$ in the regions for the coupling constant squared $(0,1)$ and $(0,30)$ in Figs (\ref{lambda5}) and (\ref{lambda6}).

In the range $(0,1)$ for the squared coupling constant there are solutions for each of the three groups in the approximate range $0.3-0.5$ but none is really close to the smaller value of $g_s^2$ or $e^2$ obtained in Eqs. (\ref{sol76885}) and (\ref{sol891745}). In the range $(0,30)$  there is a single solution acceptable for the group $SO(10)$ at $g_{10}^2\approx 20$ corresponding to the larger values for $g_s^2$ and $e^2$ in Eqs. (\ref{sol76885}) and (\ref{sol891745}). Thus our conjecture indicates that the only viable and natural candidate for a supersymmetric gauge unification and for an ultraviolet completion of the standard model up to the Planck scale (among the three main groups considered here) is $SO(10)$.

\section{Conclusions}

Calculating the beta function in higher orders for a gauge theory is by itself a very difficult and important enterprise important not only for disentangling various properties but also for revealing the phase diagram of the theory.  However for all knowledge to gain phenomenological relevance one must know either the intrinsic scale of the theory be that infrared or ultraviolet or at least the value of the coupling constant at some scale. The absence of such information is a significant gap of knowledge which the present paper aimed to fill at least partially.

In this work we proposed a conjecture valid for any gauge theory scale invariant at tree level that allowed the determination of the coupling constant at some definite scale by comparison to a known gauge theory. This is completely equivalent to finding the $\Lambda$ scale of the theory.

We showed that this conjecture worked perfectly for $U(1)_{em}$ and $QCD$, the two gauge theories for which there is phenomenological information. Based on these  and by analyzing the supersymmetric GUT groups $SU(5)$, $SO(10)$ and $E_6$ we were able to make a prediction with regard to the group that is more plausible, from the point of view of the conjecture proposed here,  to complete the standard model at higher scale. It turned out that this group was $SO(10)$.

There are many possible applications that one might consider as our conjecture applies to any gauge group, abelian or non-abelian, supersymmetric or not. However these are beyond the scope of the present work.

\end{document}